\documentclass[preprint,5p,times,twocolumn]{elsarticle}

\usepackage{graphicx}
\usepackage{epstopdf}
\usepackage{color}
\usepackage{bm}
\usepackage{amssymb}
\usepackage{amsfonts}
\usepackage{amsmath}
\usepackage{dcolumn}
\usepackage{float}
\usepackage{multirow}
\usepackage{hyperref}
\hypersetup{
    colorlinks,
    citecolor=blue,
    filecolor=blue,
    linkcolor=blue,
    urlcolor=blue
}
\journal{Computational Materials Science}
\begin{document}
\begin{frontmatter}

\title{Intrinsic strength and failure behaviors of ultra-small single-walled carbon nanotubes}

\author[a]{Nguyen Tuan Hung\corref{cor1}}
\cortext[cor1]{Corresponding author. Tel.:+81 22 795 7754; Fax: +81 22 795 6447.}
\ead{nguyen@flex.phys.tohoku.ac.jp}
\author[b,c]{Do Van Truong}
\author[b,c]{Vuong Van Thanh}
\author[a]{Riichiro Saito}

\address[a]{Department of Physics, Tohoku University, Sendai 980-8578, Japan}
\address[b]{International Institute for Computational Science and Engineering, Hanoi University of Science and Technology, Hanoi, Vietnam}
\address[c]{Department of Design of Machinery and Robot, Hanoi University of Science and Technology, Hanoi, Vietnam}

\begin{abstract}
The intrinsic mechanical strength of single-walled carbon nanotubes (SWNTs) within the diameter range of 0.3-0.8 nm has been studied based on \textit{ab initio} density functional theory calculations. In contrast to predicting ``smaller is stronger and more elastic" in nanomaterials, the strength of the SWNTs is significantly reduced when decreasing the tube diameter. The results obtained show that the Young's modulus $E$ significantly reduced in the ultra-small SWNTs with the diameter less than 0.4 nm originates from their very large curvature effect, while it is a constant of about 1.0 TPa, and independent of the diameter and chiral index for the large tube. We find that the Poisson's ratio, ideal strength and ideal strain are dependent on the diameter and chiral index. Furthermore, the relations between $E$ and ideal strength indicate that Griffith's estimate of brittle fracture could break down in the smallest (2, 2) nanotube, with the breaking strength of 15\% of $E$. Our results provide important insights into intrinsic mechanical behavior of ultra-small SWNTs under their curvature effect.
\end{abstract}


\end{frontmatter}

\section{Introduction}
\label{sec:introduction}
Due to one-dimensional (1D) structures~\cite{Saito}, the single-walled carbon nanotubes (SWNTs) are the ideal material for a variety of applications relating to tensile strain. The SWNTs and graphene are known as the strongest materials with ultrahigh axial Young's modulus of about 1.0 TPa and tensile strength approaching 100-130 GPa~\cite{Lee,Poncharal,Krishnan,Wu,Yao,Treacy,Fereidoon,Xiao}. Both experimental and theoretical studies~\cite{Wu,Ogata,Mori} showed that the diameter of the large SWNTs does not significantly affect their mechanical properties. However, the physical and mechanical properties of ultra-small SWNTs with the diameters smaller than 0.4 nm expected are different from those larger than that due to their very large curvature effect. Many efforts have been made to synthesize the ultra-small SWNTs in recent years. The smallest stable (2, 2) SWNT with a diameter of 0.3 nm observed by Zhao et al.~\cite{Zhao} could be grown inside multi-wall carbon nanotubes (MWNTs). The (2, 2) nanotube investigated by the first principle calculations is tunable between metallic and semiconducting properties by changing the Fermi level~\cite{Yin}. In fact, the ultra-small nanotubes are less stable than the large nanotubes~\cite{Qin,Wang}. However, if we fabricated nicely in some special geometries, we can measure the values of the ultra-small SWNTs. Although, many studies have focused on synthesis, physical and chemical properties of the small SWNTs~\cite{Zhao,Yin,Lortz,Sasaki,Tang,Li}, their mechanical properties have yet to be clarified. Moreover, the intrinsic mechanical properties such as Young's modulus, Poisson's ratio and ideal strength are key factors relating to the stability and lifetime of devices. For these reasons, studying the mechanical response of small SWNTs under strain should be a necessary task in order to improve the future SWNTs-based devices.

The SWNT structure is unique due to the strong bonding between the carbons ($sp^2$ hybridization of the atomic orbitals) of the curved graphene sheet, which is stronger than in diamond with $sp^3$ hybridization because of the difference in C-C bond lengths (0.142 and 0.154 nm for graphene and diamond respectively)~\cite{Saito}. The changes in the C-C bond structure such as defects, grain boundaries, chemical substitutions or curvature effects are the main causes to make changes in mechanical properties of SWNTs and graphene. The results obtained by the density function theory (DFT) and molecular dynamic (MD) calculations showed that the Young's modulus and tensile strength of SWNTs~\cite{Sammalkorpi,Wong,Zhang,Sharma} and graphene~\cite{Hao} with vacancy-related defects depend on the concentration of defects and defect characteristics. Zhang et al.~\cite{Zhang1} investigated that the grain boundaries (GBs) are significantly reduced the mechanical strength of graphene. Mortazavi et al.~\cite{Mortazavi} reported that the Young's modulus of a nitrogen doped in a graphene is almost independent of nitrogen atom concentration, but the substituted nitrogen atoms are decreased the tensile strength and ductile failure behavior of graphene. For the perfect (5, 5), (6, 3) and (8, 0) SWNTs, the tight binding (TB) and DFT calculations showed that SWNTs can reach the Young's modulus of 1.0 TPa and a maximum tensile strength of 100 GPa with no chiral dependence~\cite{Mori}. However, the critical tensile strain for breaking has a chiral dependence. The experiment~\cite{Wu} has used the optical characterization with a magnetic actuation technique to measure the stiffness of the (17, 12), (17, 10) and (18, 10) SWNTs and found that the Young's modulus is not dependent on the nanotube chiral index, and has an average value of 0.97 $\pm$ 0.16 TPa. This mechanical response is also observed in the graphene. Both the experiment and DFT calculation reported the Young's modulus of 1.0 TPa for both the zigzag and the armchair tensile strain directions~\cite{Lee,Liu}. In 1920, Griffith~\cite{Griffith} extrapolated an maximum intrinsic strength $\sigma_I$ of about $E/9$ for the fracture of brittle material, where $E$ is the Young's modulus of the material under uniaxial tension. This estimate is still valid for the brittle material in nano-scale. Both the experiment and theory showed that $\sigma_I/E$ is approximately 0.1 for the graphene and nanotubes~\cite{Lee,Liu,Ogata}. Therefore, it is interesting to investigate that the mechanical response of the large SWNTs and graphene is consistent with the small SWNTs, which are dominated by their very large curvature effect.

In this paper, we present the first-principles to investigate the structural and mechanical properties of the small SWNTs with the diameter in the range from 0.3 to 0.8 nm under uniaxial tension. The paper is organized as follows. Section \ref{sec:methodology} describes the setup of the DFT calculations and the detailed simulation procedure. Section \ref{sec:results} describes Young's modulus, Poisson's ratio, ideal strength and fracture mechanism of the SWNTs under tensile strain. Finally, section \ref{sec:conclusion} summarizes the results.

\section{Methodology}
\label{sec:methodology}
First-principle (\textit{ab initio}) simulations for tensile strains of small diameter single-walled carbon nanotubes (SWNTs) was performed. We used Quantum-ESPRESSO (QE) package~\cite{Giannozzi1} for the first-principle calculations, which is a full density functional theory~\cite{Hohenberg,Kohn} simulation package using a plane-wave basic set with pseudopotentials. The Rabe-Rappe-Kaxiras-Joannopoulos (RRKJ)~\cite{Rappe} ultrasoft pseudopotentials was used to calculate the pseudopotential plane-wave with an energy cutoff of 60 Ry for the wave function. The exchange-correlation energy was evaluated by general-gradient approximation(GGA)~\cite{PSEUDO} using the Perdew-Burke-Ernzerhof (PBE)~\cite{Perdew} function.

We examined three models of a series of small diameter single-walled carbon nanotubes: the armchair type (2, 2), (3, 3), (4, 4), (5, 5), (6, 6) SWNTs; the zigzag type (3, 0), (4, 0), (5, 0), (6, 0), (7, 0) SWNTs and the chiral type (3, 1), (3, 2), (4, 1), (4, 2), (5, 2) SWNTs, which have the diameters in the range from 0.3 to 0.8 nm. Here, the SWNT structure in our notation is denoted by a set of integers ($n, m$) which is a shorthand for the chiral vector $\textbf{C}_h = n\textbf{a}_1 + m\textbf{a}_2$, where $\textbf{a}_1$ and $\textbf{a}_2$ are the unit vectors of an unrolled graphene sheet~\cite{Saito}. The chiral vector $\textbf{C}_h$ defines the circumferential direction of the rolled-up graphene into a cylinder, giving the diameter $D_0$. The crosssectional area of a SWNT layer was calculated using the interlayer distance of a MWNT (0.34 nm)~\cite{Ge} as its thickness. Since a periodic boundary condition was applied for three dimensions in all models, the thickness of the vacuum region was set at 12 \AA\ perpendicular to the tube axis to avoid the undesirable interactions from the neighboring SWNTs. The \textbf{k}-point grids in the Brillouin-zone selected according to the Monkhorst-Pack method~\cite{Monkhorst} was $1 \times 1 \times k$, in which $k$ depends on the length of the SWNTs. 

\begin{figure}[h]
\begin{center}
\includegraphics[width=0.48\textwidth]{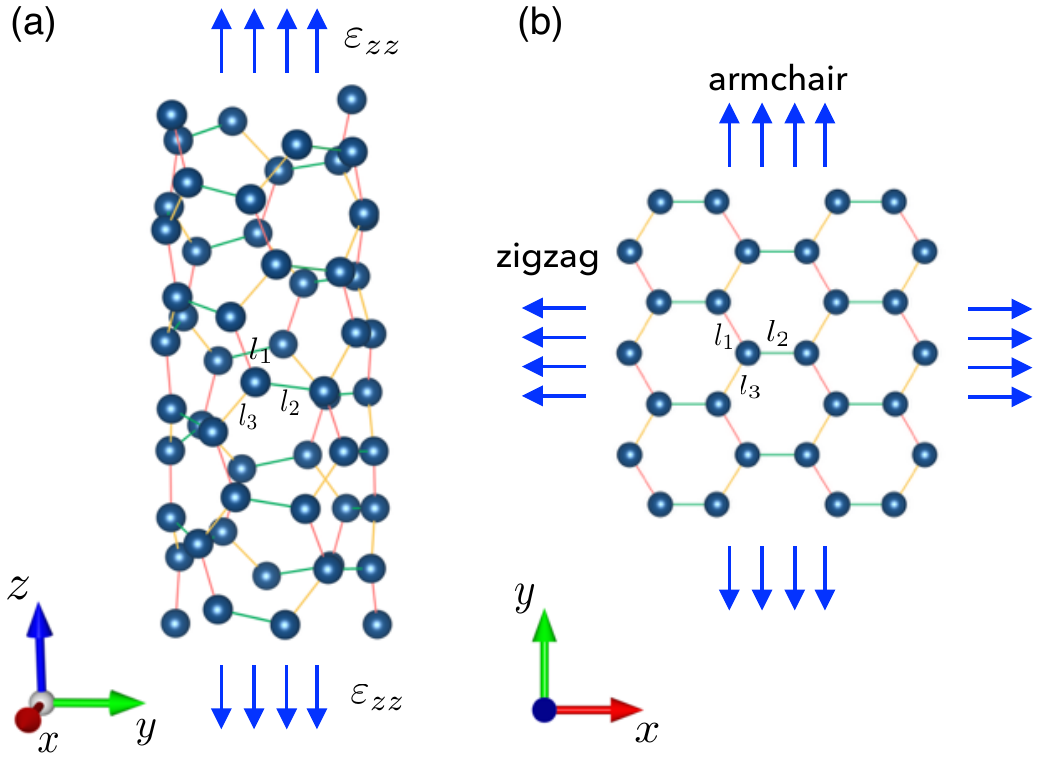}
\caption{(Color online) (a) The simulation model of small nanotubes under tensile strain in \textit{z} direction. (b) The model of graphene under tensile strain in zigzag and armchair directions. The notations of $l_1$ to $l_3$ are three bond lengths around C atomic.}
\label{fig:model}
\end{center}
\end{figure}

To simulate the effect of tensile strain in the SWNTs, first, the models were fully relaxed by using the Broyden-Fretcher-Goldfarb-Shanno (BFGS) minimization method for the atomic positions, and cell dimensions in the \textit{z} direction. These models were considered equilibrium until all the Hellmann-Feynman forces and the normal component of the stress $\sigma_{zz}$ less than $5.0 \times 10^{-4}$ Ry/a.u. and $1.0 \times 10^{-2}$ GPa, respectively. Then the loading strain was applied to the models by elongating the cell along the \textit{z} direction with an increment of 0.02. At near the fracture point, the strain was refined with a very small increment of 0.005. After each increment of the strain, the atomic structure was fully relaxed under fixed cell dimensions. Here, the tensile strain is defined as $\varepsilon_{zz}\equiv \Delta L/L_{0}$, where $L_0$ is the length of the unit cell at geometry optimization and $\Delta L$ is the increment of the length under tension [Fig. \ref{fig:model}(a)]. We also investigated the mechanical response of graphene under tensile strain in the zigzag and armchair directions to elucidate curvature effect, as shown in Fig. \ref{fig:model}(b).

\begin{table*}[t]
\caption{Equilibrium configurations including number of atom in unit cell $n$, length $L_0$, diameter $D_0$, bond length $l_{1,2.,3}$, and binding energy $E_b$ and Young's modulus $E$, Poisson's ratio $\nu$, ideal strains $\varepsilon_{I}$, and ideal strength $\sigma_{I}$ for different SWNTs.} 
\centering    
{\renewcommand{\arraystretch}{1.2}    
\begin{tabular}{c {c}*{11} c}
\hline\hline
SWNT & $n$ & $L_0$ (nm) & $D_0$ (nm) & $l_1$ (nm) & $l_2$ (nm) & $l_3$ (nm) & $E_b$ (eV) & $E$ (TPa) & $\nu$ & $\varepsilon_{I}$ & $\sigma_{I}$ (GPa) & $\sigma_{I}/E$ \\ 
\hline    
(2, 2) & 8 & 0.257 & 0.282 & 0.149 & 0.139 & 0.149 & 7.770 & 0.576 & 0.147 & 0.270 & 86.44 & 0.150 \\    
(3, 3) & 12 & 0.246 & 0.420 & 0.143 & 0.144 & 0.143 & 8.365 & 0.954 & 0.067 & 0.305 & 97.90 & 0.103 \\ 
(4, 4) & 16 & 0.246 & 0.553 & 0.143 & 0.143 & 0.143 & 8.578 & 0.965 & 0.108 & 0.325 & 101.44 & 0.105 \\
(5, 5) & 20 & 0.246 & 0.686 & 0.143 & 0.143 & 0.143 & 8.676 & 0.981 & 0.104 & 0.320 & 103.80 & 0.106\\
(6, 6) & 24 & 0.246 & 0.821 & 0.142 & 0.142 & 0.142 & 8.729 & 0.978 & 0.093 & 0.310 & 105.44 & 0.108\\
\hline
(3, 0) & 12 & 0.421 & 0.263 & 0.149 & 0.141 & 0.149 & 7.682 & 0.765 & 0.100 & 0.235 & 102.59 & 0.131 \\
(4, 0) & 16 & 0.421 & 0.337 & 0.148 & 0.138 & 0.148 & 8.090 & 0.842 & 0.124 & 0.215 & 102.58 & 0.120 \\ 
(5, 0) & 20 & 0.426 & 0.407 & 0.145 & 0.141 & 0.145 & 8.343 & 0.914 & 0.112 & 0.185 & 89.53 & 0.096 \\
(6, 0) & 24 & 0.426 & 0.483 & 0.144 & 0.141 & 0.144 & 8.493 & 0.936 & 0.106 & 0.205 & 102.65 & 0.108 \\
(7, 0) & 28 & 0.426 & 0.559 & 0.143 & 0.141 & 0.143 & 8.586 & 0.997 & 0.085 & 0.180 & 102.29 & 0.101 \\
\hline  
(3, 1) & 52 & 1.536 & 0.303 & 0.145 & 0.145 & 0.145 & 7.894 & 0.817 & 0.087 & 0.175 & 83.58 & 0.102 \\
(3, 2) & 76 & 1.865 & 0.356 & 0.144 & 0.143 & 0.144 & 8.201 & 0.887 & 0.093 & 0.215 & 81.56 & 0.092 \\ 
(4, 1) & 28 & 0.649 & 0.357 & 0.145 & 0.144 & 0.142 & 8.253 & 0.927 & 0.112 & 0.210 & 93.27 & 0.101 \\
(4, 2) & 56 & 1.127 & 0.428 & 0.143 & 0.143 & 0.144 & 8.390 & 0.935 & 0.093 & 0.235 & 77.36 & 0.083 \\
(5, 2) & 52 & 0.887 & 0.500 & 0.143 & 0.143 & 0.143 & 8.517 & 0.929 & 0.119 & 0.210 & 95.82 & 0.103 \\
\hline\hline
\end{tabular}
}
\label{table:nanotubes}    
\end{table*}

\section{Results and discussions}
\label{sec:results}
\begin{figure}[h]
\begin{center}
\includegraphics[width=0.45\textwidth]{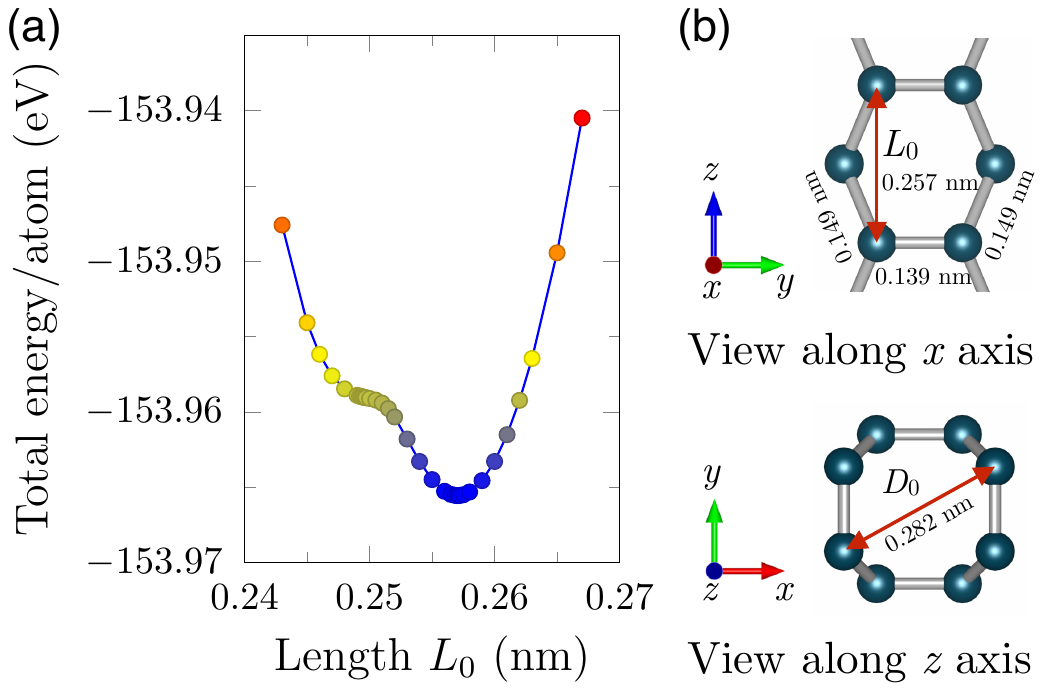}
\caption{(Color online) (a) Total energy per atom of the (2,2) SWNT plotted as a function of the tube length. (b) Unit cell of the (2,2) SWNT showing the length $L_0$, the diameter $D_0$, the C-C bond perpendicular the tube axis, and the that along to the tube axis.}
\label{fig:(2,2)tube}
\end{center}
\end{figure}

The scanning of the potential energy surface (PES) is performed for the smallest (2, 2) SWNT to search for the ground state with the length $L_0$ ranging from 0.24 to 0.27 nm by calculating the total energy per atom with different lengths. Fig.~\ref{fig:(2,2)tube}(a) shows the observed minimum of the total energy at $L_0=0.257$ nm. In Fig.~\ref{fig:(2,2)tube}(b), we present the illustrations of atomic structures for the (2, 2) SWNT at the ground state, where $L_0 = 0.257$ nm, $D_0=0.282$ nm, and $l_1 = l_3 = 0.149$ nm ($l_2=0.139$ nm) represent the length, the diameter, and the C--C bond length parallel to (perpendicular to) the tube axis, respectively. The same optimization is performed for the other nanotubes. The equilibrium configurations are listed in Table \ref{table:nanotubes} for all SWNTs in this study. The results show that the C-C bond lengths of the small nanotubes tend to be longer than those of the large nanotubes. In the calculation of the binding energy, we take the energy of an isolated C atom ($E_{\rm C}$) as the reference energies, with $E_{\rm tot}$ being the total energy of the system containing $n$ C atoms in the unit cell. The binding energy per C atom, $E_b=E_{\rm tot}-nE_{\rm C}$, is summarized in Table~\ref{table:nanotubes}. It is found that the binding energy becomes larger when the diameter of the SWNTs increases. In other words, the large SWNTs are more stable than the small SWNTs.

\begin{figure}[h]
\begin{center}
\includegraphics[width=0.4\textwidth]{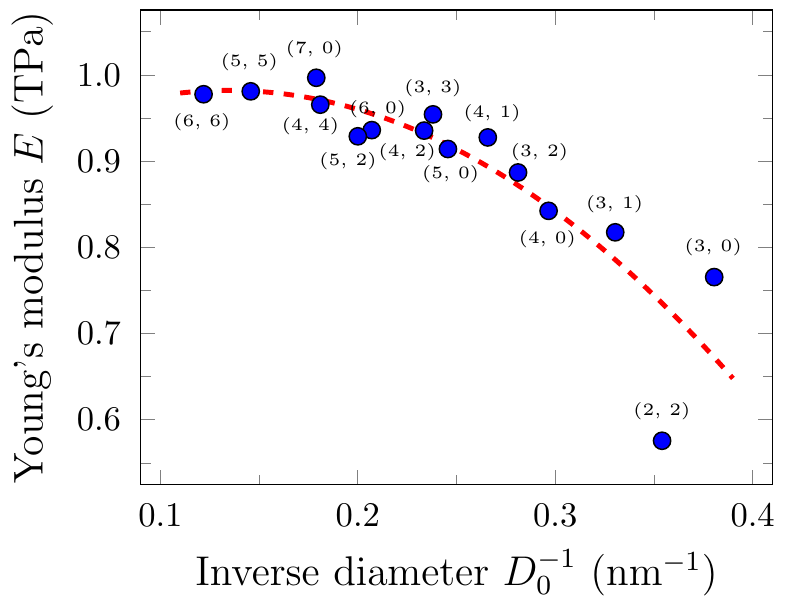}
\caption{(Color online) Young's modulus of the single-walled carbon nanotubes as a function of inverse of tube diameter. The dash line is fitted by a second-order polynomial.}
\label{fig:young}
\end{center}
\end{figure}

The most basic mechanical property of the SWNTs is the Young's modulus $E$, which is defined as
\begin{equation}
  \label{eq:young}
  E = \left. {{1}\over{V_0}}{{\partial^2U}\over{\partial \varepsilon_{zz}^2}} \right\vert_{\varepsilon_{zz}=0} 
\end{equation}
where $U$ is the strain energy and $\varepsilon_{zz}$ is the uniaxial strain. We applied the small strains ($\pm 0.005, \pm 0.01, \pm 0.015, \pm 0.02$), which stay in the harmonic regime. Here, the nominal plate thickness $d_0$ assumed was independent of $\varepsilon_{zz}$. The nanotubes with the diameter $D_0>d_0$ and $D_0 \leq d_0$ were considered as the hollow and solid cylinders, respectively. The volume at equilibrium $V_0$ is defined as
\begin{equation}
  \label{eq:volume}
  V_0=\left\{
    \begin{array}{l l}
    \pi L_0 D_0 d_0 & \quad \text{if} \quad D_0 > d_0 \\
    \pi L_0 (D_0/2+d_0/2)^2 & \quad \text{if} \quad D_0 \leq d_0 \\
    \end{array} \right.
\end{equation}
From the point of view of elasticity theory, it is well-recognized that the value of $E$ is related to $d_0$ of the tube. The wall thickness is considered as the interlayer spacing of graphite and multi-walled carbon nanotubes in nature based on the van der Waals interactions~\cite{Liu}, in which $d_0$ assumed is independent of the strain. Both the experimental and the theoretical studies~\cite{Lee,Liu} have used the constant thickness of 0.334 nm to calculate the mechanical properties of the graphene. In this study, $d_0$ of 0.34 nm, which was observed in experimental images of MWNT~\cite{Ge}, is used to estimate $E$ of the SWNTs. As show in Table \ref{table:nanotubes}, the geometric structures of the (2, 2), (3, 0), and (3, 1) nanotubes are the solid cylinders with $D_0 < d_0$. While the remaining nanotubes are the hollow cylinders with $D_0 > d_0$.

Figure~\ref{fig:young} shows the obtained Young's modulus $E$ of the nanotubes, in which they are a constant of around 1 TPa and independent of the inverse of tube diameter $1/D_0$ and the chiral index ($n, m$) for the large SWNTs with $D_0 > 0.4$ nm. These results are in good agreement with both the previous experiment and theory ones~\cite{Wu,Xiao,Ogata,Mori,Yang}. For the small SWNTs with $D_0 \leq 0.4$ nm, $E$ investigated depends on the tube diameter [Fig. \ref{fig:young}]. It is well-known that in nanomaterials, the smaller they are the stronger and more elastic~\cite{Zhu,Hung}, however the results obtained show that the rule is broken when $D_0 \leq 0.4$ nm. $E$ of the (2, 2) nanotube is decreased about 41\% comparing with the (6, 6) nanotube. The significant reduction is due to the curvature of the small nanotubes. In Fig. \ref{fig:model}, since the $l_2$ sigma bond is perpendicular to the tube axis, the intrinsic strength of the armchair nanotubes originates from the $l_1$ and $l_3$ sigma bonds. The $l_1$ and $l_3$ in the (3, 3), (4, 4), (5, 5) and (6, 6) nanotubes are very close to the value 0.142 nm of the graphene, and their $E$ are similar to the experimental value of about 1 TPa~\cite{Lee, Wu}, while affected by the curvature, the $l_1$ and $l_3$ in the (2, 2) nanotube are longer than those and its $E$ is only 0.587 TPa [Table \ref{table:nanotubes}]. For the zigzag nanotubes, because the $l_2$ sigma bond is parallel with the tube axis, the tensile strength is mainly characterized by the $l_2$ bond. However, when the diameter of the tubes is smaller than 0.4 nm (for the (3, 0) and (4, 0) nanotubes), the strength also depends on the $l_1$ and $l_3$ bonds [Table \ref{table:nanotubes}]. For the chiral nanotubes, the tensile strength investigated is governed by not only the $l_1$, $l_2$ and $l_3$ sigma bonds, and the angles between the bonds and tube axis (for the (3, 1), (3, 2), (4, 1), (4, 2), (5, 2) nanotubes) but also the curvature of the tube (for the (3, 1) and (3, 2) nanotubes).

\begin{figure}[h]
\begin{center}
\includegraphics[width=0.4\textwidth]{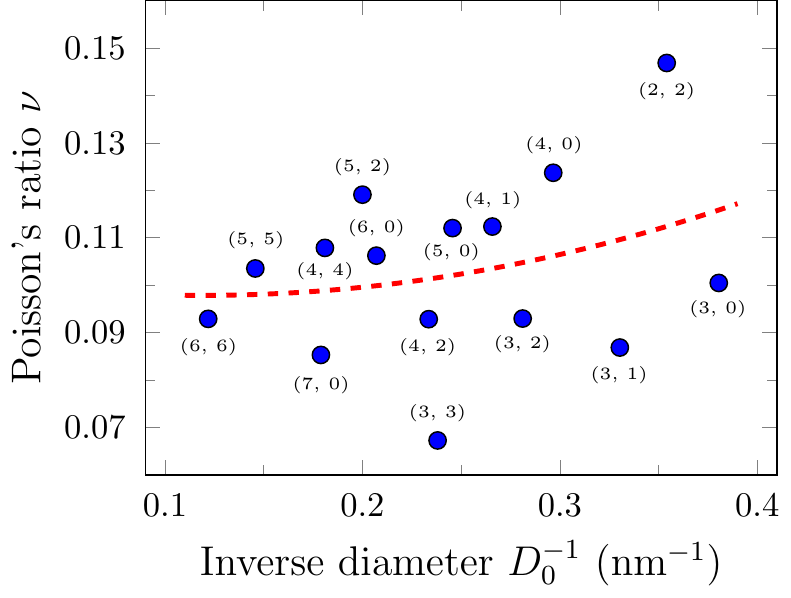}
\caption{(Color online) Poisson's ratio of the single-walled carbon nanotubes as a function of inverse of tube diameter. The dash line is fitted by a second-order polynomial.}
\label{fig:poisson}
\end{center}
\end{figure}

Figure \ref{fig:poisson} shows the obtained Poisson's ratio $\nu$ of the SWNTs as a function of the inverse of tube diameter. Here, $\nu$ is given as:
\begin{equation}
  \label{eq:poisson}
  \nu\equiv-\dfrac{L_{0}}{\Delta L}\dfrac{\Delta D}{D_{0}+d_0}
\end{equation}
The value of $\nu$ is from 0.07 to 0.15 and has a downward tendency with the large SWNTs. For the small ones, the Poisson's ratio is large compared with that of the larger nanotubes, especially for the (2, 2) nanotube. Since the thickness of wall of tube is correlated to the Poisson's ratio, we estimated $\nu$ for an expanded diameter $D_0+d_0$ [Eq.~\ref{eq:poisson}] instead of the tube diameter, $D_0$, as the previous studies~\citep{Portal,Yakobson}. Therefore, our results are smaller than the values obtained by Portal et al.~\citep{Portal} with DFT calculations ($\nu$ = 0.14), and Yakobson et al.~\citep{Yakobson} using Tersoff-Brenner potentials ($\nu$ = 0.19). 
\begin{figure}[h!]
\begin{center}
\includegraphics[width=0.4\textwidth]{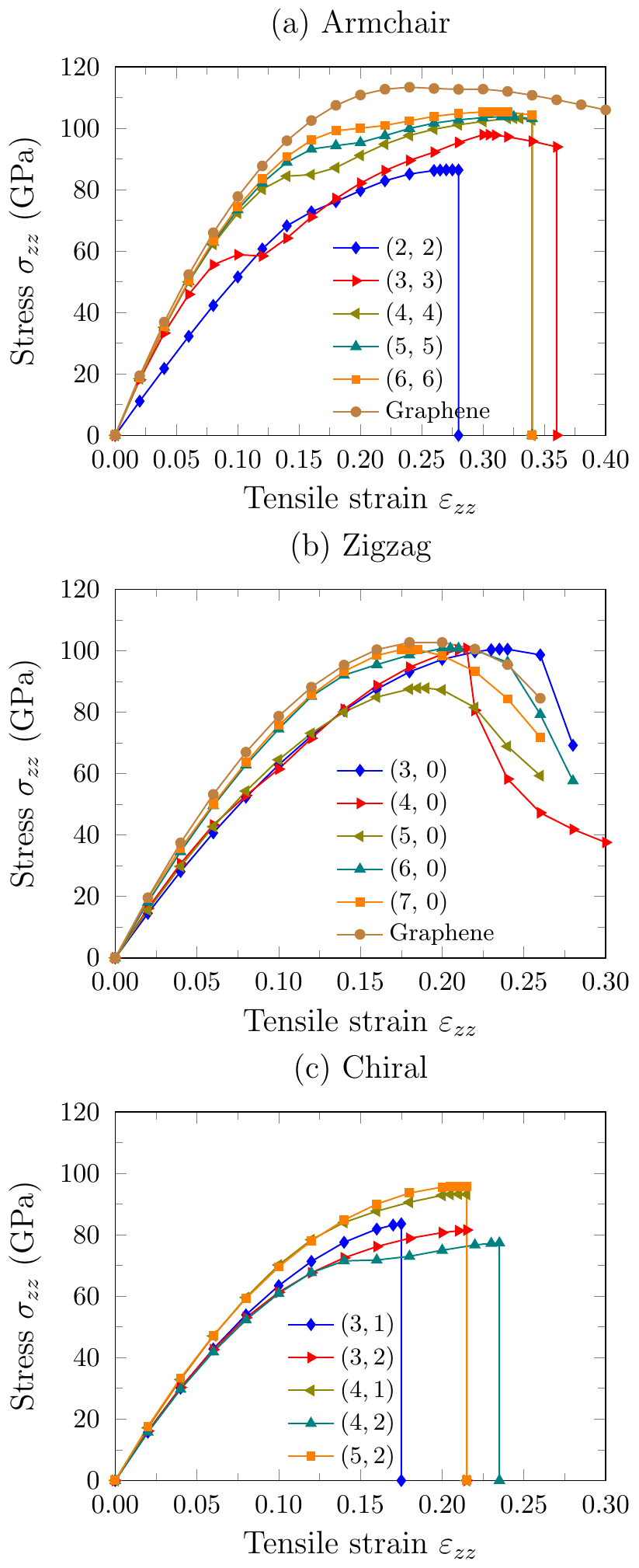}
\caption{(Color online) Tensile stress along axial direction of different armchair (a), zigzag (b) and chiral (c) nanotubes plotted as a function of strain.}
\label{fig:stress}
\end{center}
\end{figure}

Figure \ref{fig:stress} shows the stress--strain curves of the armchair, zigzag, and chiral nanotubes. The stress computed from QE package~\cite{Giannozzi1} is automatically evaluated over the entire supercell volume $V_{\rm cell}$. Therefore, we need to rescale the supercell stress by $V_{\rm cell}/V_0$ to obtain the stress of the SWNTs. The relationship between $V_0$ and $d_0$ in Eq.~\ref{eq:volume} shows that the stress is inversely proportional to the constant thickness. That means that, in general, the stress becomes larger when the magnitude of $d_0$ decreases, and vice versa. For the armchair nanotubes [Fig.~\ref{fig:stress}(a)], the ideal strength $\sigma_{I}$ (maximum tensile strength) reaches about 100 GPa at an ideal tensile strain $\varepsilon_{I}$ of around 0.30, which are consistent with the earlier DFT estimate of 114.6 GPa at $\varepsilon_{I}$ of 0.295 for the armchair nanotube~\cite{Ogata}, and in agreement with the experimental observations ($150\pm 45$ GPa) for a defect-free MWCNT using the transmission electron microscope (TEM)~\cite{Demczyk}. In the armchair direction [Fig. \ref{fig:stress}(a)], the graphene is somewhat stronger than the nanotubes, with $\sigma_{I}$ of 113.4 GPa. The ideal strength of graphene between 110 and 130 GPa have been also predicted by both the experiment~\cite{Lee} and the DFT calculation~\cite{Liu}. For the zigzag nanotubes, the ideal strength of SWNTs is similar to that of the armchair nanotubes and graphene in the zigzag direction with $\sigma_{I}$ of around 100 GPa [Fig. \ref{fig:stress} (b)]. However, $\varepsilon_{I}$ of around 0.20 is smaller than that of the armchair nanotubes. These ideal strength and ideal strain are compatible with the earlier DFT calculation of 107.4 GPa and 0.208 for an zigzag nanotube, respectively~\cite{Ogata}. For the chiral nanotubes, $\sigma_{I}$ and $\varepsilon_{I}$ are found approximately 90 GPa and 0.20, respectively. The results obtained above show that $\sigma_{I}$ and $\varepsilon_{I}$ depend not only on the tube diameter, but also on the tube chiral index. The ideal strength and the ideal strain evaluated from the stress--strain relationship [Fig. \ref{fig:stress}] are listed in Table \ref{table:nanotubes}. 

The relationship between the Young's modulus and the ideal strength has been known as Griffith's estimate of brittle fracture~\cite{Griffith}. Due to $\sigma_{I}$ from 90 to 100 GPa and $E\sim 0.1$ TPa for the large nanotubes, the ratio $\sigma_{I}/E$ has the value between 0.09 and 0.10 [Table \ref{table:nanotubes}]. The breaking strength of nanotube is reaching 10\% of its Young's modulus. This upper theoretical limit has been predicted by both the experiment and the theory for graphene~\cite{Lee,Liu} and SWNTs~\cite{Ogata}. However, the small nanotubes show that $E$ is significantly decreased while $\sigma_{I}$ is in the range from 83 to 102 GPa [Table \ref{table:nanotubes}]. In particular, the ratio $\sigma_{I}/E$ about 15\% of the smallest (2, 2) nanotube can lead to broken Griffith's estimate.

\section{Conclusion}
\label{sec:conclusion}
In summary, \textit{ab initio} density-functional theory calculations with the general gradient approximation have been carried out to investigate the intrinsic mechanical strength of the single-walled carbon nanotubes with the different chiralities and diameters under tensile strain. The results obtained reveals that the intrinsic strength in the nanotubes originates from the sigma bonds. The atomic structures and the bond lengths of small SWNTs (with diameter $<$ 0.4 nm) are significantly changed under their very large curvature effect. The strength of the small SWNTs is significantly weaker than the large ones. This is in contrast with the phenomenon ``smaller is stronger and more elastic" in nanomaterials~\cite{Zhu,Hung}. For the large SWNTs, the Young's modulus $E\sim 1.0$ TPa is independent of the diameter and the chiral index. These results are in good agreement with the previous experimental and the theoretical studies~\cite{Wu,Demczyk,Ogata,Lee,Liu}. The Poisson's ratio $\nu$ has a noticeable downward trend with the large SWNTs. For the small nanotubes, $\nu$ is large compared with that of the larger nanotubes, especially for the (2, 2) nanotube. The ideal strength obtained from 90 to 100 GPa and the ideal strain from 0.20 to 0.30 depends on the diameter and the chiral index. Furthermore, the relationships between the Young's modulus and the ideal strength indicate that Griffith's estimate of brittle fracture could break down in the smallest (2, 2) nanotube, with the breaking strength of 15\% of $E$.

\section{Acknowledgements}
N.T.H. acknowledges the support of the Interdepartmental Doctoral Degree Program for Multi-dimensional Materials Science Leaders of Tohoku University, R.S. acknowledges MEXT Grants Nos. 25107005 and 25286005, and D.V.T. and V.V.T. acknowledges NAFOSTED No. 107.02.2012.20.

\end{document}